# RTNN: The new parallel machine in Zaragoza[*]

A.J. van der Sijs

*Departamento de Física Teórica, Universidad de Zaragoza,
Facultad de Ciencias, 50009 Zaragoza, Spain*

I report on the development of RTNN, a parallel computer designed as a $4^4$ hypercube of 256 T9000 transputer nodes, each with 8 MB memory. The peak performance of the machine is expected to be 2.5 Gflops.

## 1 Introduction

The RTNN project is a collaboration of the Universidad de Zaragoza, the Universidad Autónoma de Madrid and the Universidad Complutense de Madrid. Its objective is to build a 256-node parallel computer with the topology of a $4^4$ hypercube, where each node contains a T9000 transputer with 8 MB of memory and 10 Mflops peak. It is the successor of RTN, a 100 Mflops parallel machine based on the T805 transputer built a couple of years ago with a budget of US$ 50 000, which has been fully operational since November 1991. RTN stands for Reconfigurable Transputer Network, and the additional N in RTNN (or RTN$^2$) for New.

The RTN project, initiated by the same Spanish Universities which now form the RTNN collaboration, was motivated by the need of large-scale computing facilities in Spain, and the lack of funds for these purposes. The Italian APE project, in which some Spanish researchers had participated, was an additional stimulus; in fact, the APE group formally joined the RTN collaboration at a later stage.

The new RTNN project, which was approved in 1993 (at least, funding for half the machine which was estimated to cost US$ 400 000 was allocated), has been delayed substantially by problems in the development and production of the INMOS T9000 processor. At the moment of writing, testing of the prototype RTNN base modules is finished, and the design of the base boards is in its final stage. The machine may be operational by the end of 1995 or early 1996.

---

## 2  RTN

The "old" RTN consists of 64 T805 transputers, of the INMOS T800 series. These transputers are 20 MHz RISC processors with 4 kB of internal (cache) memory and an FPU with a peak speed of 1.5 Mflops, leading to an overall peak performance of about 100 Mflops. Each transputer has its own memory (1 MB of DRAM).

The 64 transputers are distributed over 8 boards containing 8 transputers each. The 8 transputers on each board, labelled from 0 to 7, are connected (in hardware) in a circle, and the 8 boards are usually connected in a circle as well, leading to an $8 \times 8$ torus topology. The R for Reconfigurability refers to the possibility to change the connections *between* boards, not *within* boards. This transputer network is furthermore connected to a host (a workstation) via a control board.

Each T805 transputer has four bi-directional "links" for information exchange. Two of them are used for the direct connections to the neighbouring transputers on the same board. The third link of each of the eight transputers on a board is connected to an INMOS C004 "Cross Link" chip on the same board, while the fourth link of each transputer is connected to the Cross Link on the preceding board. This Cross Link is a kind of telephone central, in charge of the communications between links. All it does is connecting its (incoming and outgoing) links in pairs. These connections can be changed by sending the Cross Link a special configuration command. This is how (partial) reconfigurability is achieved.

Note that in the two-dimensional torus configuration, all communication links of all the transputers are occupied, and no communication with the outside world (*i.e.*, with the host) is possible. This problem is circumvented by a subtle use of the Event link (an additional channel for warning messages) of one of the number-zero transputers connected to the control board. Through this link the host tells the transputer network to "close the torus" and start running the program, and upon completion (or crashing!) the system tells the host that the program is finished and that the torus can be "opened" again, for transmission of results to the host.

Information exchange is simpler in the new RTNN machine, thanks to improved communication skills of the T9000 transputer and the use of a "more intelligent" Cross link. As we shall see, this makes it possible to configure the T9000's as a four-dimensional hypercube, although each of them has only four communication links available.

As mentioned earlier, RTN has been operational since late 1991. Since then, it has been used for the numerical investigation of topics as wide-ranging as
- Phase diagram of the U(1)-Higgs model [1];
- Instantons in SU(2) gauge theory with twisted boundary conditions [2];
- Vacuum structure of two-dimensional $\phi^4$ and $\phi^6$ models [3];
- Four-dimensional Ising spin glasses [4];
- Geophysical fracture models [5];
- Critical behaviour of random walks [6];

The list of citations is by no means exhaustive, only a few key references have been given.

## 3 RTNN and the T9000 transputer

For the RTNN project the INMOS T9000 transputer was chosen, because of its good communication abilities via the four links, and because of previous experience with INMOS transputers in the RTN project. The T9000, originally announced for 1993, is a superscalar processor with a 16 kB instruction and data cache, and 20 to 30 MHz speed options. The transputer itself generates these clock frequencies from a single 5 MHz input signal. At 30 MHz, it is advertised to have a peak performance of 120 MIPS ($> 50$ sustained) and 10 MFLOPS ($> 7$ sustained). (The Floating Point Unit can handle 64-bit words.) The plan to make available a version of the T9000 running at 50 MHz appears to have been abandoned.

The T9000 has 4 communication links, but with a higher transmission capacity (40 MB/s total bi-directional link bandwidth) than the T800 series and a more advanced communication protocol ("Daisy Chains", see Sect. 5, and so-called "Virtual Channel Processing" which will not be discussed here). A further difference with the T800 series is the presence of two control links which can be used for (re)configuration, so the complicated tricks with the Event links, used in the operation of RTN, can be avoided.

Unfortunately, the construction of RTNN has been severely delayed by problems in the development and production of the T9000 chip. A 20 MHz version, containing minor bugs, has been available since 1994. It is now hoped that a reliable version of the T9000, running at 30 MHz, will become available in the course of 1995.

## 4 Topology of RTNN

The 256 T9000 nodes of RTNN will be arranged as a $4^4$ hypercube, built as a $2^4$ hypercube of boards, with each board itself laid out as a $2^4$ hypercube of processor units. If in the end funds turn out to be sufficient for only half the machine, the number of boards will be eight instead and the topology $4^3 \times 2$. Although this picture may suggest fixed connections between the transputers, this is not the case. Neither have any (periodic) boundary conditions been imposed, which would turn the hypercube into a four-dimensional torus. In fact, no two transputers are directly connected, as will be discussed shortly.

First I would like to insert a brief intermezzo about why a $4d$ hypercubic topology is preferred in comparison with a square or $3d$ cubic configuration. This has to do with minimizing the amount of communication between transputers. Suppose we want to carry out a lattice Monte Carlo study of a system on an $L^4$ lattice, with $N = n^4$ transputers at our disposal. Assume that $L^4/N$ is a large number. In the $4d$ hypercubic configuration, each transputer is responsible for a block of $(L/n)^4$ lattice sites. Information from the $(L/n)^3$ sites on each of the eight three-dimensional faces of this block has to be exchanged with the eight neighbouring transputers. This "transferred boundary" amounts to $8\,(L/n)^3 = 8\,L^3/N^{3/4}$ units of information per transputer per unit time step. It is easily checked that for three and two-dimensional topologies, with the obvious allocation of lattice sites to transputers, the transferred boundary is $6\,L^3/N^{2/3}$ and $4\,L^3/N^{1/2}$, respectively. (All numbers up to $\mathcal{O}(N^{1/d}/L)$ corrections.) Fig. 1 is a log-log plot of the transferred boundary as a function of the number of transputers. One sees that for $N \geq 32$ the transferred boundary is minimal in the $4d$ hypercubic configuration. At the same time, this optimizes the use of transputer memory, given that it is convenient for each transputer to allocate memory for the data received from the neighbouring processors, so that each piece of data required from the neighbours needs to be requested and transmitted only once.

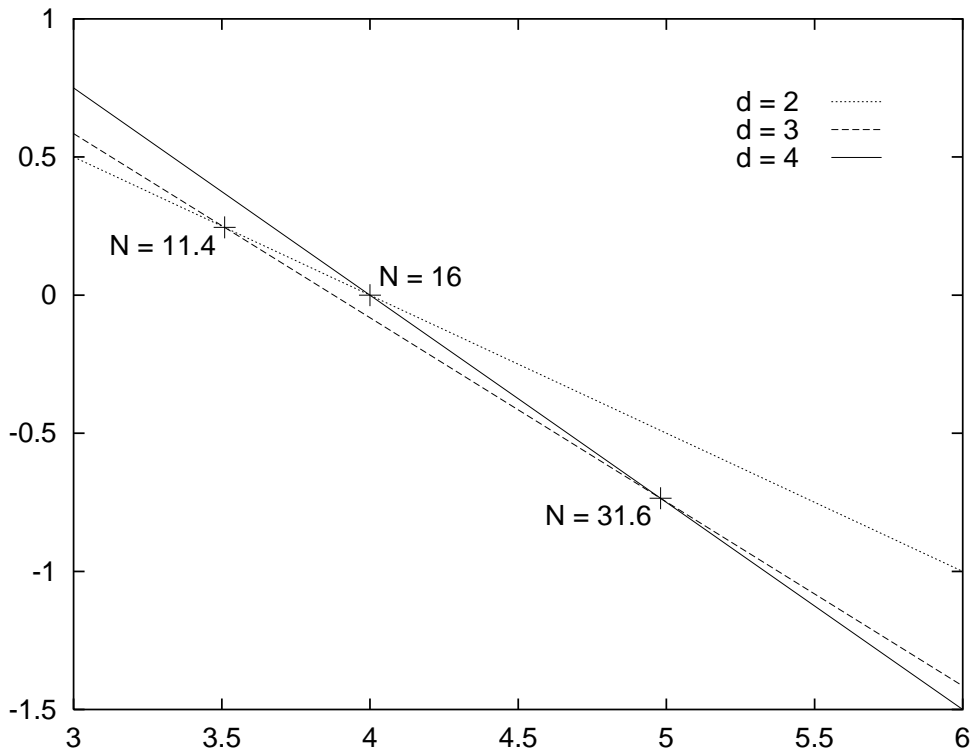

Fig. 1. Transferred boundary as a function of the number $N$ of transputers, in the limit $L \to \infty$ (i.e., ignoring higher order corrections). Plotted is $^2\log(2d/N^{(d-1)/d})$ versus $^2\log N$, for $d = 2, 3, 4$.

## 5  Communication in RTNN

The crucial communication problem for each transputer in the hypercubic configuration is obviously how to serve its 8 nearest neighbours with only 4 communication links available. Since I find this one of the most interesting aspects of the design of RTNN, I would like to discuss in some detail how it is dealt with.

The solution is provided by using INMOS C104 Cross Link chips, in the following way. The four communication links of each transputer are labelled $x, y, z, t$. The $x$-link takes care of communications with the neighbouring transputers in *both the forward and backward* $x$-directions of the $4^4$ hypercube, and similarly for the $y, z$ and $t$-links. Let us concentrate on the $x$-links. Recall that there are 16 transputers on a board, arranged as a $2^4$ hypercube. They may be labelled by coordinates $x, y, z, t$, each taking the value 0 or 1. For eight of the 16 transputers, having $x$-coordinate equal to 0, the forward neighbour is on the same board, and the backward neighbour on the preceding board, while for those lying in the $3d$ hyperface at $x = 1$ the backward neighbour is on the same board and the forward neighbour on the next one. (In fact, in the actual situation of a $2^4$ hypercube (a four-torus) of boards the preceding board and the next board are one and the same!) The 16 $x$-links of the transputers on the board are fed into a Cross Link located on the same board, which we may call the $x$-Cross Link. This occupies 16 of the Cross Link's total of 32 links. Of its remaining 16 links, 8 are connected to the $x$-Cross Link on the preceding board and 8 to the one on the next board. All these connections are physically fixed, in hardware.

What remains is the freedom to "configure" the $x$-Cross Link. When an $x = 0$ transputer needs to communicate with its forward neighbour, which is the $x = 1$ transputer on its own board having equal $y, z$ and $t$ coordinates, the $x$-Cross Link on that board connects the $x$-links of both directly, according to the "Daisy Chain" protocol described below. When this $x = 0$ transputer needs to communicate in the backward direction, on the other hand, the $x$-Cross Links are "configured" such that each transputer is connected to its neighbour on the neighbouring board.

The $y$ and $z$ directions are dealt with in the same way, with $y$- and $z$-Cross Links on each board. If the same procedure was applied for the $t$ direction, all the links of all the transputers and Cross Links would be occupied, and communication with the host would become a problem. For this reason, one has opted for a solution using two mutually connected Cross Links for communication in the $t$-direction, one taking care of the transputers in the $t = 0$ hyperplane and another for those at $t = 1$. Thus a total of five Cross Links is required on each board.

It should be stressed that the feasibility of this approach depends crucially on the flexibility offered by the communication protocol used by the T9000 transputers and the INMOS C104 Cross Link chips. In this "Daisy Chain" protocol, each chunk of information is preceded by a header which specifies its (final) destination or contains configuration information for a Cross Link. Reconfiguration can be done "along the way". This is to be contrasted with the situation in RTN, where each reconfiguration requires separate commands, in which case the present method would involve an unacceptable loss of efficiency.

## 6  Layout of RTNN

After discussing the motivation for a four-dimensional hypercubic topology and explaining how the associated communication tasks are dealt with, it is time to sketch how this hypercubic topology is physically realized in RTNN.

The basic unit in RTNN is the base module, see Fig. 2. It measures 8.7 cm. $\times$ 8.7 cm., and contains one T9000 transputer, five memory chips of 1 M $\times$ 16 bits (= 2 MB) DRAM each, an Error Detection And Correction (EDAC) chip, another (ALTERA) chip in charge of the "logistics", and a connector to the base board.

Four of the five memory chips are used as 1 M of 64-bit words of memory, providing the transputer with 8 MB of memory, the fifth is used (half) to store check bits. Each 64-bit word written in memory by the T9000 is fed through the EDAC first, which generates 8 control bits using the Hamming code. When this word is read from memory again, it goes back through the EDAC, which checks its contents against the control bits. Single errors are detected and corrected by the EDAC, multiple errors are detected only and cause a warning signal.

The base modules will be placed on base boards, measuring 40 cm. by 30 cm., which in turn are attached to the back plane via a connector. Each board is designed to contain 8 base modules at the front (see Fig. 3), and 8 at the back, with the Cross Links, dealing with the communication, in the middle (three at the front, two at the back). In addition, the front of the board contains a 5 MHz oscillator providing the signal from which the transputers on the board generate their own 20 or 30 MHz clock frequency; a reset facility; three chips for the control logic; and a number of buffers used both for amplifying signals to be sent out of the board via the connector and for blocking (possibly unintended) unwelcome signals (error messages, resets) emitted by other boards. The latter point is related to the issue of scalability and segmentation, see Sect. 8. Finally, each board boasts a set of LEDs and digital error counters, for visualization of the machine's performance (see Sect. 7).

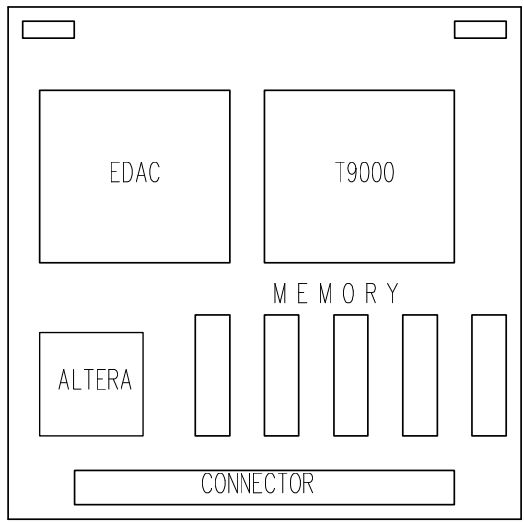

Fig. 2. Base module of RTNN, measuring 8.7 cm. × 8.7 cm.

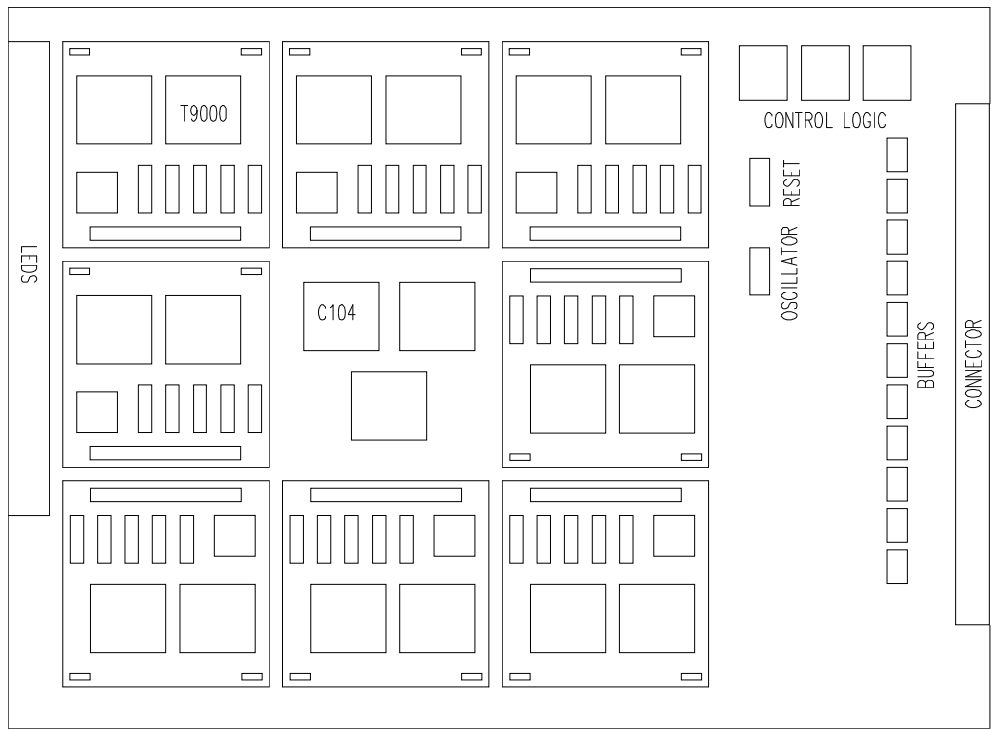

Fig. 3. Front of the base board of RTNN, measuring 40 cm. × 30 cm. There are also eight base modules at the back, as well as two C104's.

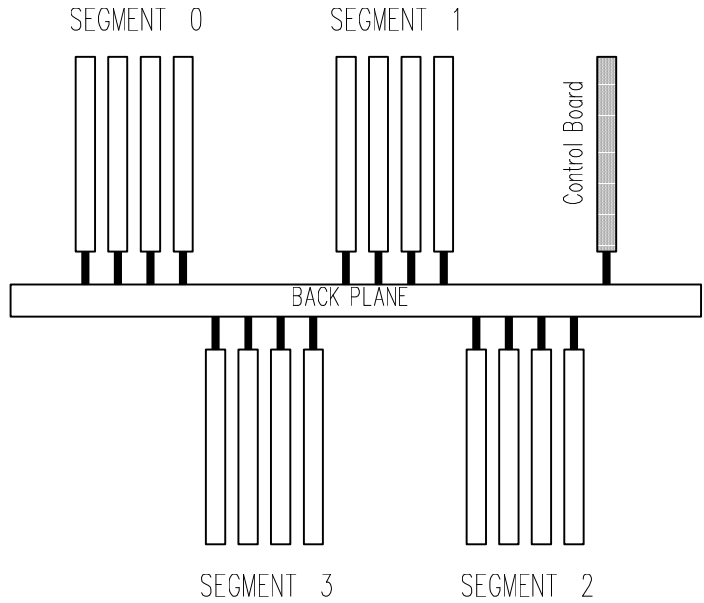

Fig. 4. Location of the base boards and control board in the back plane. The distance between two boards within a segment is about 5 cms. (recall that both the front and the back of the boards contain base modules).

The base boards are inserted into the back-plane (in groups of four, see Fig. 4), which is responsible for providing the necessary connections among the boards.

## 7 Visualization

For a "home-made" machine, it is not a bad idea to build in a few devices to monitor how well (if at all) its various parts are working. For this purpose, each base module is connected to a set of 2 LEDs located at the edge of the board, whose four states off-off, off-on, on-off, on-on correspond to idle (no activity), memory access, multiple error, and single error, respectively. The mapping between the status of the module and the LED illumination has been devised in such a way that one can in fact obtain some useful information from these cute little lights (given that a LED continuously switching on and off with a frequency of the order of MHz is always observed as "on" by the human eye).

In addition, each base module sends error signals from its EDAC chip to a 2-digit decimal error counter. (Why 2-digit? It might be too optimistic to expect fewer than 10 errors in a reasonable amount of time, while more than 100 errors in a short period of time would probably make the whole machine useless anyway.)

Finally, a very simple but useful indicator of the activity of the machine is the power being consumed during operation.

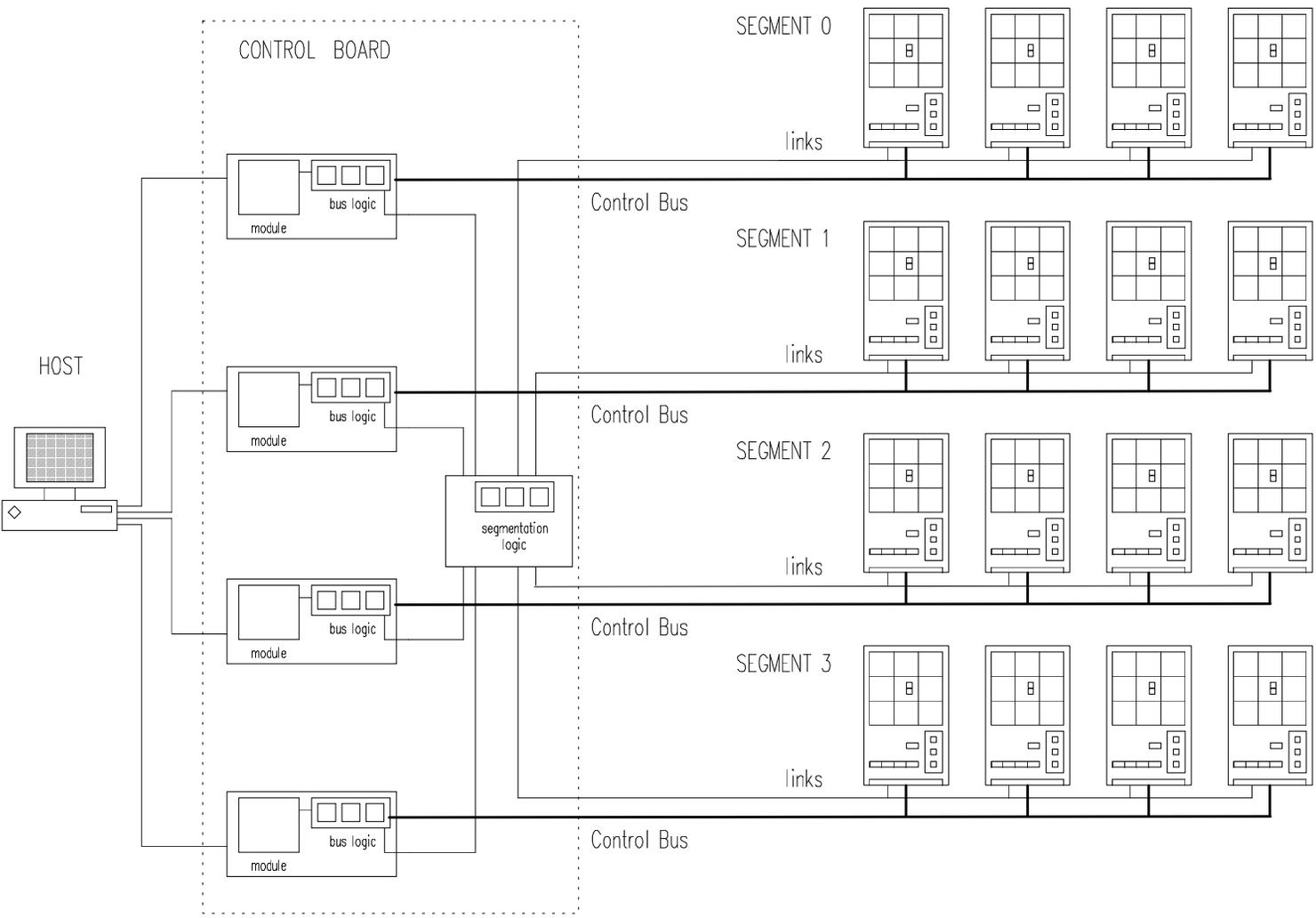

Fig. 5. Schematic view of RTNN, showing clearly the segmentation of the machine.

## 8  Segmentation

One of the properties of RTNN is scalability, related with the possibility of segmentation, dividing the machine into a number of smaller units, each of which can be assigned to a different user.

First of all, the base module consisting of one T9000 with its memory, EDAC and logic chips, can be operated on its own, when connected to a host. This is how prototype base modules are being tested at present.

At the next level, a base board containing 16 base modules together with the five Cross Links for communication can be used as an independent unit. As mentioned earlier, the set of buffers on the board can be used as a barrier against unwelcome signals arriving from other boards.

The most important level of segmentation, however, is the partition of the machine into autonomous segments consisting of four boards (64 processors), the $x$-$y$-planes of the $2^4$ hypercube formed by the complete set of boards. This is sketched in Fig. 5. Each segment has its own control unit, which is a slightly adapted base module on the control board. Control information to and from all the four boards in the segment is transmitted from and to the control unit over a control bus containing 11 lines, two of which identify which of the 4 boards is the destination or sender of the message. This control bus is completely separate from the channels exchanging the data from the transputer links.

The host can now assign a block of one, two or four control units, with their corresponding segments of four boards, to a single user. Any remaining segments can be assigned to other users. One of the tasks of the control board is now to keep the Cross Links of the base boards configured in such a way that no signals can be sent out to segments not belonging to a different user, *i.e.*, signals in the $z$ and $t$-directions in case the user has one segment, and signals in the $t$-direction only if the user has two segments. The segmentation of the control logic should furthermore provide the security that error signals produced by user $A$ do not force an emergency stop in a program of user $B$ running in a different part of the machine.

## 9  Cooling

A less interesting but nevertheless non-negligible aspect of the construction of RTNN is cooling. A rough estimate of the dissipation of the machine, obtained by simply adding up the power consumed by the various components, yields a number of the order of 5 kW: Without adequate cooling RTNN would be an oven, quite suitable for baking tortillas de patata at lunch time. The definitive solution remains to be decided, but is not yet clear that a ventilator blowing air between the boards will suffice.

## 10  Status and Outlook

The construction of RTNN has been severely slowed down by delays in the development and production of the T9000 transputer. In the meantime, other cheap and fast processors have appeared on the market which have made the T9000 far less competitive, and besides, attractively priced commercial parallel machines have become available. Because of its good communication properties, thanks to its four data links, and because of previous experience with the T805 transputer, the RTNN collaboration has nevertheless remained loyal to the T9000 so far. Apart from this, the money allocated to the RTNN project cannot easily be spent in a way too different from the original proposal.

At present a 20 MHz version of the T9000 chip is available and used for testing the base module of RTNN. The final specifications of the chip which will ultimately be used are not yet known, and neither is the final number of T9000's, 128 or 256, that can be purchased.

As for the present status of the project, the base module has successfully passed the test phase, the electronic design of the base boards is about finished, and the design of the control board and the back plane is on its way. The hope is to finish construction of the machine by the end of the year. Physics may then be done in 1996.

## Acknowledgements

It is a pleasure to thank the organizers, in particular A. Nakamura and his "family" of students, for a very enjoyable workshop.

Furthermore, I am grateful to C. López Ullod, A. Cruz, D. Iñiguez and A. Tarancón from whom I learned most of what I know about RTN(N).

The RTNN project is funded by the High Energy Physics Program of the General Secretariat of the National Plan for Research and Development (Spain), project number AEN 93-0604.

This work is supported by EC contracts CHRX-CT92-0051 and ERBCHBICT-941067, by DGICYT Spain, project number AEN 94-218 and by Acción Integrada Hispano-Francesa HF94-150B.

## References

[1] The RTN collaboration, J.L. Alonso et al., Nucl. Phys. B 405 (1993) 574; Phys. Lett. B 296 (1992) 154.
[2] The RTN collaboration, M. García Pérez et al., Phys. Lett. B 305 (1993) 366.
[3] J. Casahorrán, J.C. Ciria, J.G. Esteve and A. Tarancón, Phys. Lett. B 303 (1993) 305.
[4] D. Badoni, J.C. Ciria, G. Parisi, F. Ritort, J. Pech and J.J. Ruiz-Lorenzo, Europhys. Lett. 21 (1993) 495.
[5] J.B. Gómez, D. Iñiguez and A.F. Pacheco, Phys. Rev. Lett. 71 (1993) 380.
[6] I. Campos and A. Tarancón, Phys. Rev. E 50 (1994) 91.